\PassOptionsToPackage{sort&compress}{natbib}
\documentclass[5p]{elsarticle}
\makeatletter 
\def\ps@pprintTitle{%
 \let\@oddhead\@empty
 \let\@evenhead\@empty
 \let\@evenfoot\@oddfoot} 
\makeatother
\usepackage[utf8]{inputenc}
\usepackage[T1]{fontenc}

\newtheorem{theorem}{Theorem}
\newtheorem{example}{Example}

\newdefinition{definition}{Definition}
 
\newproof{proof}{Proof}

\usepackage{pdflscape}
\usepackage{mathptmx} 
\usepackage{helvet}
\usepackage{courier}
\usepackage{type1cm}
\usepackage{url}
\usepackage{makeidx}
\usepackage[bottom]{footmisc}
\usepackage[colorlinks = true,linkcolor = blue,urlcolor = blue,citecolor = blue,anchorcolor = blue]{hyperref}
\usepackage{amsmath,amsfonts,amssymb,amscd}
\usepackage{graphicx}
\usepackage{epstopdf}
\usepackage{pifont}
\usepackage{subcaption}
\usepackage{color}
\usepackage{multirow}
\usepackage{multicol}
\usepackage[boxed,ruled,commentsnumbered,vlined]{algorithm2e}
\usepackage{tikz}
\usepackage{pgfplots}
\usepackage{float}
\usepackage{fancyhdr}
\usepackage{booktabs}
\usepgfplotslibrary{groupplots}
\usetikzlibrary{pgfplots.groupplots}
\usetikzlibrary{plotmarks}
\usetikzlibrary{matrix}
\usepackage{array}
\newcolumntype{x}[1]{>{\centering\arraybackslash\hspace{0pt}}p{#1}}
\usepackage{rotating}
\usepackage[misc]{ifsym}
\makeindex

\usepackage{natbib}
\usepackage{CJKutf8}

\begin{document}

\begin{frontmatter}

\title{Mining compact high utility sequential patterns}

\author{Tai Dinh}
\address{The Kyoto Colleges of Graduate Studies for Informatics}
\cortext[corresponding]{Corresponding Author}
\ead{t\_dinh@kcg.ac.jp}
\author{Philippe Fournier-Viger}
\address{College of Computer Science and Software Engineering, Shenzhen University}
\author{Huynh Van Hong}
\address{Ho Chi Minh University of Natural Resources and Environment}

\begin{abstract}
High utility sequential pattern mining (HUSPM) aims to mine all patterns that yield a high utility (profit) in a sequence dataset. HUSPM is useful for several applications such as market basket analysis, marketing, and website clickstream analysis. In these applications, users may also consider high utility patterns frequently appearing in the dataset to obtain more fruitful information. However, this task is high computation since algorithms may generate a combinatorial explosive number of candidates that may be redundant or of low importance. To reduce complexity and obtain a compact set of frequent high utility sequential patterns (FHUSPs), this paper proposes an algorithm named CHUSP for mining closed frequent high utility sequential patterns (CHUSPs). Such patterns keep a concise representation while preserving the same expressive power of the complete set of FHUSPs. The proposed algorithm relies on a CHUS data structure to maintain information during mining. It uses three pruning strategies to eliminate early low-utility and non-frequent patterns, thereby reducing the search space. An extensive experimental evaluation was performed on six real-life datasets to evaluate the performance of CHUSP in terms of execution time, memory usage, and the number of generated patterns. Experimental results show that CHUSP can efficiently discover the compact set of CHUSPs under different user-defined thresholds.
\end{abstract}


\end{frontmatter}

\section{Introduction}
Frequent high utility sequential pattern mining (FHUSPM) finds sequential patterns with high utility and frequently appear in sequence datasets. Such patterns appear commonly in various real-life applications such as market basket analysis, web- site clickstream analysis, customer behavior analysis, and stock market analysis. In market basket analysis, when analyzing customer transactions, a retail store manager may be interested in finding the high utility patterns that appear regularly and have a high sale volume. Detecting these purchase patterns is useful for understanding customers’ behavior and thus adopting effective sales and marketing strategies. For example, high-end electronic devices and jewelry may generate more profit than many daily-life products. However, they may be sold infrequently, and their sales volumes may greatly fluctuate. Suppose retailers know that some products yield a high profit and are frequently purchased; they can change business strategies for these items to increase sales and improve inventory management. In marketing, marketers want to know some sets of products frequently sold with high revenue. They can better understand customers’ preferences and then design efficient marketing strategies. In website clickstream analysis, the number of clicks or time spent on each web page or user interface (UI) element can be viewed as the quantities of items in sequences. Thus, administrators can discover the elements where users spend most of their time. Based on that, administrators can improve functions and UI to suit these important behaviors better.

Although the problem of HUSPM and its extensions have been studied in several previous  \cite{yin2012uspan,wang2016efficiently,le2018pure,gan2020proum,dinh2015novel,le2018efficient,huynh2022multi}, these algorithms discover a full set of HUSPs requiring exponential complexity. Therefore, in this paper, we extend the concept of closed patterns from frequent sequential pattern mining \cite{wang2004bide} for HUSPM. A closed (frequent) high utility sequential pattern (CHUSP) is a HUSP having no proper super-sequences that are HUSPs and appear in the same number of sequences. Such patterns are also meaningful for real-life applications since they are the largest FHUSPs common to groups of customers. Detecting the largest sets of items yielding high profit and frequently sold supports sellers to understand better what customers need, adapt their business and marketing strategies, and improve their services. There is a work \cite{truong2019fmaxclohusm} focusing on this topic in literature. However, the computational complexity of this algorithm is still high. In addition, the experimental evaluation was conducted on small-scale datasets which a few differences in characteristics. Last, this work did not provide the application accompanying its proposed algorithms.

The above observations motivated the design of an efficient algorithm that can mine CHUSPs. Generally, we highlighted the major contributions and innovations of this paper as follows:
\begin{itemize}
\item[-] We proposed an efficient pattern-growth-based algorithm named CHUSP to discover the set of CHUSPs interesting for some tasks. CHUSP mines the patterns from the dataset in a divide-and-conquer approach. It first derives the set of size-1 quantitative sequences, and for each sequence $p$, it derives $p$’s conditional (or projected) dataset by partitioning it and recursively mining the projected dataset. An innovation of the CHUSP is that the algorithm checks the ``closed'' property of the generated pattern at each round of the mining process. Thanks to this property, at the end of the mining process, we obtain a small set of CHUSPs. The algorithm uses two pruning strategies to eliminate early low-utility and non-frequent patterns. Thus, the algorithm achieves good performances on large-scale datasets.
\item[-] An extensive experiment was conducted on real datasets to evaluate the performance of CHUSP in terms of runtime, memory usage, and the number of generated patterns. Experimental results show that CHUSP can efficiently discover all CHUSPs. In addition, its performance is independent of the datasets' characteristics as long as they contain utility information, i.e., it can work on both quantitative transaction or quantitative sequence datasets.
\item[-] We provide the application of CHUSP. The application can be used for any dataset if its format matches the input requirement. 
\end{itemize}
The rest of this paper is organized as follows. Section 2 reviews related work; section 3 introduces the preliminaries; section 4 describes the proposed CHUSP algorithm; section 5 shows a comparative experiment; section 6 concludes and outlines the direction for future work.
\section{Related work}
High utility sequential patterns mining aims to find all sequential patterns with a utility greater than or equal to a minimum utility threshold $minUtil$ in a sequence dataset. HUSPM is quite challenging as the utility measure is neither monotone nor anti-monotone, unlike the support measure traditionally used in SPM. Numerous algorithms have been proposed for HUSPM, and its extension \cite{yin2012uspan,wang2016efficiently,le2018pure,gan2020proum,dinh2015novel,le2018efficient,huynh2022multi,fournier2016mining,dinh2017mining,quang2016mhhusp,huynh2021hiding,dinh2018efficient,quang2016approach,dinh2020k,huynh2021mining,dinh2019survey,fournier2020mining,xie2022efficient}. A thorough survey of HUSPM can be found at this work \cite{truong2019survey}. Yin et al. \cite{yin2012uspan} proposed an algorithm named USpan for HUSPM. This algorithm builds a lexicographic q-sequence tree (LQS-Tree) to maintain all generated sequences during the mining process. In addition, it uses two concatenation mechanisms: I-Concatenation and S-Concatenation, in combination with two pruning strategies: width and depth pruning. Wang et al. \cite{wang2016efficiently} proposed an algorithm named HUS-Span. The algorithm uses a utility-chain structure to represent the search space of HUSPM. It also introduces two tight utility upper bounds: {prefix extension utility} (PEU) and {reduced sequence utility} (RSU), as well as two companion pruning strategies to identify HUSPs. The experimental evaluation showed that HUS-Span outperforms USpan in terms of execution time. The reason is that by using PEU and RSU, HUS-Span can generate fewer candidates than USpan. 

Le et al. \cite{le2018pure} proposed two algorithms named AHUS and AHUS-P. The algorithms use a pure array structure (PAS) to represent sequences. This data structure is very compact and contains sufficient information on sequences. Thus, it can reduce memory usage and effectively support the mining process. Moreover, the two algorithms use two upper bounds to prune the search space. AHUS-P uses a parallel mining strategy to discover patterns concurrently by sharing the search space with multiple processors. Each processor independently performs its mining task and does not wait for other tasks. AHUS-P is more efficient than the serial AHUS algorithm for large-scale datasets. Lin et al.~\cite{lin2020efficient} proposed a sequence-utility (SU)-Chain algorithm for HUSPM. A lexicographic enumeration (LE)-tree is used in the algorithm to represent the search space for promising candidates. The projecting approach is used to accelerate the progress of generating promising candidates. In addition, multiple pruning strategies are used to identify information not relevant to the mining progress.

For frequent high utility sequential pattern mining, Gupta et al. \cite{gupta2022hufti} proposed a hybrid pattern growth-based algorithm named HUFTI-SPM to mine sequential patterns satisfying both frequency and utility thresholds. It uses support–utility table to maintain information on support and utility at various time intervals. It uses sequence support as the downward closure property to reduce the search space. Ni et al. \cite{ni2022frequent} proposed an algorithm named FHUSOM to mine the architecture design requirements from the operational scenario data. The algorithm uses a data structure called FHUDS to keep all patterns and combines four pruning strategies called SWU, PEU, RSU, and MFP to reduce the search space. The algorithm supports the design of an integrated multi-platform mission system (MPMS) architect and is efficient in the process of integrated architecture design.

For closed high utility sequential pattern mining, Truong et al.~\cite{truong2019fmaxclohusm} proposed an algorithm named FMaxCloHUSM to mine the set of frequent maximal and closed high utility sequences. The algorithm uses the width and depth pruning strategies to remove low utility sequences and a novel local pruning strategy named LPCHUS to remove non-closed and non-maximal high utility sequences. FMaxCloHUSM uses a data structure called SIDUL to represent the dataset in a vertical format and calculate utility information of sequences and their extensions.

\section{Preliminaries}
Given a set of $m$ distinct items $I=\{{i_1},{i_2},\dots,{i_m}\}$. A quantitative item (q-item) is a pair of the form $(i,q)$ where $i \in I$ and $q$ is a positive number representing how many units of this item were purchased (internal utility). The quantity of a q-item $i$ in $s$ is denoted as $q(i,s)$. Each item ${i_k} \in I$ $(1 \le k \le m)$ is associated with a weight denoted as $p({i_k})$ representing the unit profit or importance (external utility) of ${i_k}$. A quantitative itemset (q-itemset) $X= [({i_1},{q_1})({i_2},{q_2})...({i_k},{q_k})] $ is a set of one or more q-items where $({i_j},{q_j})$ is a q-item $(1 \le j \le k)$. In the following, brackets are omitted for brevity if a q-itemset contains only one q-item. In addition, without loss of generality, assume that q-items in a q-itemset are sorted according to the lexicographical order (e.g., $a$ $\prec$ $b$ $\prec$ $c$ $\prec$ $d$ $\prec$ $e$ $\prec$ $f$ $\prec$ $g$). A quantitative sequence (q-sequence) $s$ is an ordered list of q-itemsets $s = \langle {I_1}{I_2}...{I_l}\rangle $ where ${I_j} (1 \le j \le l)$ is a q-itemset. A quantitative sequence dataset is a set of $n$ q-sequences $SDB$= $\{s_1,s_2,\dots,s_n\}$, where each sequence $s_{sid} \in S$ $(1 \le sid \le n)$ is a subset of $I$, and $sid$ is its unique identifier.
\begin{example} \normalfont
\normalfont
Table \ref{externalUtility} shows the items and their respective unit profits appearing in an online retail store. 
In this example, the external utility of each item $a$, $b$, $c$, $d$, $e$, $f$, and $g$ are 2, 5, 3, 4, 6, 1, and 7, respectively. Table \ref{sequencedataset} shows five shopping q-sequences with quantities, having the sequence identifiers ($sid$) 1 to 5 (denoted ${s_1}$ to ${s_5}$). Each q-sequence comprises one or more transactions (q-itemsets). Each transaction in a q-sequence has a unique transaction identifier ${tid}$, and consists of one or many q-items. 
The q-sequence $s_4$ contains three q-itemsets $[(b,1)$$c(1)$$(e,2)$$(g,5)]$, $[(b,2)$$(c,1)$$(e,2)]$ and $[(a,3)$$(b,2)$$(e,4)$$(f,2)]$ in which the internal utility of q-item $e$ in the first, second and third q-itemsets are $2$, $4$ and $2$, respectively. We use the notation ${i_{tid}}$ to refer to the occurrence of the item $i$ in the $tid$-th transactions of a q-sequence. In ${s_2}$, the notation ${c_1}$ means that the q-item $c$ appears in the first q-itemset of ${s_2}$, that is $(c,2)$, while ${c_3}$ represents $(c,1)$ in the third q-itemset of ${s_2}$, and ${c_1} \prec {c_3}$ in ${s_2}$.
\end{example}

\begin{table}[h]
\renewcommand{\arraystretch}{1.0}
\centering
\caption{External utility values }
\begin{tabular}{|l|c|c|c|c|c|c|c|}
\hline
item & $a$ & $b$ & $c$ & $d$ & $e$ & $f$ & $g$ \\
\hline
unit profit & 2 & 5 & 3 & 4 & 6 & 1 & 7 \\ 
\hline
\end{tabular}
\label{externalUtility} 
\end{table}
\begin{table}[h]
\renewcommand{\arraystretch}{1.0}
\centering
\caption {A sequence dataset}
\begin{tabular}{|c|c|l|c|c|}
\hline
sid	& tid & transactions & tu & su \\ 
\hline \hline
\multirow{4}{*}{1}	&	$1$ & 	$(a,5)(c,2)(g,5)$ & $51$ & 	\multirow{4}{*}{108} \\\cline{2-4}
					&	$2$ & 	$(a,3)(b,1)(c,3)(f,2)$ & $22$ &	\\\cline{2-4}
					&	$3$ & 	$(b,3)(d,2)(e,2)$ & $35$ &	\\\hline
\hline
\multirow{3}{*}{2}	&	$1$ & 	$(c,2)(e,1) $ & $12$ & 	\multirow{3}{*}{110} \\\cline{2-4}
					&	$2$ & 	$(a,2)(b,2)(f,5)$	& $19$ &	\\\cline{2-4}
					&	$3$ & 	$(b,2)(c,1)(e,4)(g,6)$ & $79$ &	\\\hline
\hline
\multirow{4}{*}{3}	&	$1$ & 	$(a,1)(b,1)(e,3)$ 	& $25$ & \multirow{4}{*}{91} \\\cline{2-4}
					&	$2$ & 	$(c,3)(d,2)(g,3)$ 		& $38$ &	\\\cline{2-4}
					&	$3$ & 	$(b,2)(e,1)$ 		& $16$ &	\\\cline{2-4}
					&	$4$ & 	$(d,3)$			& $12$ &	\\\hline
\hline
\multirow{3}{*}{4}	&	$1$ & 	$(b,1)(c,1)(e,2)(g,5)$ 	& $55$ & \multirow{3}{*}{122} \\\cline{2-4}
					&	$2$ & 	$(a,3)(b,2)(e,4)(f,2)$ 	& $42$ &	\\\cline{2-4}
					&	$3$ & 	$(b,2)(c,1)(e,2)$ 	& $25$ &		\\\hline				
\hline
				$5$	&	$1$ & 	$(a,4)(d,2)(f,2)(g,10)$ & $88$ & $88$ \\\hline
\end{tabular}
\label{sequencedataset} 
\end{table}

\begin{definition} [The size and length of a q-sequence]
The size of $s$ is the number of q-itemsets it contains. The length of $s$ is the number of q-items in $s$. In other words, $s$ is called k-q-sequence if and only if there are k q-items in $s$, i.e. $\left|s\right|= k$, where $\left|s\right| = \sum_{I_j \subseteq s}{\left|I_j\right|}$ and $\left|I_j\right|$ is the total number of q-items in the q-itemset $I_j$. For example, the size and length of $s_4$ in Table \ref{sequencedataset} are 3 and 11, respectively.
\end{definition}
\begin{definition} [q-itemset containment]
\label{sequencecontaining}
Let ${X_a}$= $[({i_{a_1}}$, ${q_{a_1}})$ $({i_{a_2}}$, ${q_{a_2}})$ \dots $({i_{a_m}}$, ${q_{a_m}})] $ and ${X_b}$= $[({i_{b_1}}$,${q_{b_1}})$ $({i_{b_2}}$,${q_{b_2}})$ \dots $({i_{b_{m'}}}$,${q_{b_{m'}}})]$ be two q-itemsets, where ${i_{a_k}} \in I$ $(1 \le k \le m)$ and ${i_{b_{k'}}} \in I$ $(1 \le k' \le m')$. If there exist positive integers $1\le j_1 \le j_2 \le$ $\dots$ $\le j_m \le m'$, such that ${i_{a_1}} = {i_{b_{j_1}}} \wedge {q_{a_1}} = {q_{b_{j_1}}}$, ${i_{a_2}} = {i_{b_{j_2}}} \wedge {q_{a_2}} = {q_{b_{j_2}}}$, $\dots$, ${i_{a_m}} = {i_{b_{j_m}}} \wedge {q_{a_m}} = {q_{b_{j_m}}}$ then ${X_b}$ is said to contain ${X_a}$, denoted as ${X_a} \subseteq {X_b}$. For example, q-itemset $[(a,1)(b,1)(e,3)]$ in $s_3$ contains $(a,1)$, $(b,1)$, $(e,3)$, $[(a,1)(b,1)]$, $[(a,1)(e,3)]$, $[(b,1)(e,3)]$, $[(a,1)(b,1)(e,3)]$.
\end{definition}

\begin{definition} [q-subsequence]
\label{subsequence}
Given q-sequences $A$= $\langle$ $A_1$ $A_2$ $\dots$ $A_n$ $\rangle$ and $B$=$\langle$ $B_1$ $B_2$ $\dots$ $B_{n'}$ $\rangle$ $(n \leq n')$, where $A_\alpha$, $B_\beta$ are q-itemsets $(1 \le \alpha \le n$, $1 \le \beta\le n')$. If there exists positive integers $1\le$ $j_1$ $\le$ $j_2$ $\le$ $\dots$ $\le$ $j_n$ $\le$ $n'$, such that $A_1\subseteq B_{j_{1}}$, $A_2\subseteq B_{j_{2}}$, $\dots$, $A_n\subseteq B_{j_{n}}$, then $A$ is a q-subsequence of $B$ and $B$ is a q-supersequence of $A$, denoted as $A \subseteq B$. For example, $\langle [(a,5)(c,2)(g,5) ]\rangle$ and $\langle [(a,3)(b,1)(c,3)(f,2)]\rangle$ are two q-subsequences of $s_1$.
\end{definition}
\begin{definition} [Utility of a q-sequence]
\label{sequenceUtility}
The utility of an $(i,q)$ in $s$ is denoted and defined as $u(i, q) = p(i) \times q(i)$. The utility of a q-itemset $X$ in $s$ is denoted and defined as $u(X) = \sum\limits_{k = 1}^m {u({i_k},{q_k})} $. The utility of $s$ is denoted and defined as $u\left( s \right) = \sum\limits_{j = 1}^n {u({X_j})}$. 
\end{definition}
\begin{example} \normalfont
The utility of $g$ in $s_1$ (i.e. $g_1$) is $u(g, 5) = 7 \times 5 = 35$. The utility of $[(a,5)(c,2)(g,5)]$ in $s_1$ is $u([(a,5)(c,2)(g,5)])$ = $u(a, 5)$ + $u(c, 2)$ + $u(g, 5)$ = $2 \times 5 + 3 \times 2 + 7\times 5 = 51$. The utility of $s_1$ is $u(s_1)$ = $u([(a,5)(c,2)(g,5)])$ + $u([(a,3)(b,1)(c,3)(f,2)])$ + $u([(b,3)(d,2)(e,2)])$ = $51 + 22 + 35 = 108$.
\end{example}

\begin{definition} [Utility matrix]
\label{utilitymatrixdef}
A utility matrix of $s$ is $m \times n$ matrix, where $m$ and $n$ are the number of q-items and q-itemsets (transactions) in $s$, respectively. The element at the position $(k,j)$ $(0 \le k < m$, $0 \le j < n)$ of the utility matrix stores the utility $u(i_k,q)$ of the q-item $(i_k,q)$ in the q-itemset $j$. Table \ref{utilitymatrix} shows the utility matrix of $s_3$ for the sequence dataset $SDB$ depicted in Table \ref{sequencedataset}.
\end{definition}
\begin{table}[h]
\centering
\caption{The utility matrix of $s_3$}
\begin{tabular}{|c|c|c|c|c|}
\hline
item & $tid_1$ & $tid_2$ & $tid_3$ & $tid_4$ \\
\hline
a & 2 & 0 & 0 & 0 \\ 
\hline
b & 5 & 0 & 10 & 0 \\ 
\hline
c & 0 & 9 & 0 & 0 \\ 
\hline
d & 0 & 8 & 0 & 12 \\ 
\hline
e & 18 & 0 & 6 & 0 \\ 
\hline
g & 0 & 21 & 0 & 0 \\ 
\hline
\end{tabular}
\label{utilitymatrix} 
\end{table}

\begin{table}[h]
\centering
\caption{The remaining utility matrix of $s_3$}
\begin{tabular}{|c|c|c|c|c|}
\hline
item & $tid_1$ & $tid_2$ & $tid_3$ & $tid_4$ \\
\hline
a & 89 & 0 & 0 & 0 \\ 
\hline
b & 84 & 0 & 18 & 0\\ 
\hline
c & 0 & 57 & 0 & 0\\ 
\hline
d & 0 & 49 & 0 & 0\\ 
\hline
e & 66 & 0 & 12 & 0\\ 
\hline
g & 0 & 28 & 0 & 0\\ 
\hline
\end{tabular}
\label{remainingmatrix}
\end{table}
\begin{definition} [Remaining utility]
\label{remaining utility}
Given $s$ = $\langle {X_1}{X_2}...{X_n}\rangle $ where ${X_k}$=$[({i_{k_1}},{q_{k_1}})$ $({i_{k_2}},{q_{k_2}})$ \dots $({i_{k_m}},{q_{k_m}})]$ is a q-itemset of $s$. The remaining utility of q-item ${i_{k_m}}$ in $s$ is denoted and defined as $ru(i_{k_m},s)$ = $\sum \limits_{i' \in s \wedge i_{k_m} \prec i'} { u(i')} $. For example, the values $ru(a_1,s_3)$, $ru(b_1,s_3)$ and $ru(b_3,s_3)$ are respectively equal to $89$, $84$ and $18$.
\end{definition}
\begin{definition} [Remaining utility matrix]
\label{remaining utility matrix}
A remaining utility matrix of $s$ is $m \times n$ matrix, where $m$ and $n$ are the number of q-items and q-itemsets (transactions) in $s$. The element at the position $(k,j)(0 \le k < m$, $0 \le j < n)$ of the remaining utility matrix stores the $ru(i_k,q)$ of q-item $(i_k,q)$ in q-itemset $j$. For example, 
Table \ref{remainingmatrix} shows the remaining utility matrix of $s_3$ of Table \ref{sequencedataset}.
\end{definition}
\begin{definition} [Matching]
\label{matching}
Given $s = \langle ({i_1},{q_1})({i_2},{q_2})...({i_n},{q_n})\rangle$ and a sequence $t = \langle {t_1}{t_2}...{t_m}\rangle$, $s$ is said to match $t$ if and only if $n=m$ and ${i_k} = {t_k}$ for $1 \le k \le n$, denoted as $t \sim s$.
\end{definition}
\begin{example} \normalfont
Sequence $\langle(acg)(abcf)(bde)\rangle$ matches $s_{1}$. Note that because of quantities, two q-items may be considered different, although they contain the same item. Hence there could be multiple q-subsequences of a q-sequence matching a given sequence. For instance, sequence $\langle(e)\rangle$ matches respectively the q-subsequence $\langle (e,3)\rangle$ and $\langle (e,1)\rangle$ in the first and third q-itemsets of $s_{3}$. Sequence $\langle [ac]\rangle$ matches both the q-subsequences $\langle[(a,5)(c,2)]\rangle$ and $\langle[(a,3)(c,3)]\rangle$ of $s_{1}$.
\end{example}
\begin{definition} [Ending q-item maximum utility]
\label{ending q-item utility}
Given a sequence $s$ = $\langle$${x_1}{x_2}\dots{x_n}\rangle$ where ${x_j} (1 \le j \le n)$ is a q-itemset and a sequence $t = \langle {t_1}{t_2}...{t_m}\rangle$. If any q-subsequence ${s_a} = \langle {x_{a_1}}{x_{a_2}} \dots {x_{a_m}}\rangle$ $({s_a} \subseteq s$ and ${s_a} \sim t)$ where ${x_{a_m}}=[({i_{a_1}},{q_{a_1}})({i_{a_2}},{q_{a_2}}) \dots ({i_{a_m}},{q_{a_m}})]$, then $({i_{a_m}},{q_{a_m}})$ is called the ending q-item of sequence $t$ in $s$. The ending q-item maximum utility of a sequence $t$ in $s$ is denoted and defined as $u(t,i,s) = \max \{ u(s')|s'\sim t \wedge s' \subseteq s \wedge i \in s'\} $.
\end{definition}
\begin{example} \normalfont
The ending q-items of $t = \langle bd \rangle $ in $s_3$ are $d_2$, $d_4$ and their ending q-item maximum utility are respectively $u(\langle$ $bd \rangle$,$d_2$,${s_3})$ = $\max(13)$ = $13$, $u(\langle bd \rangle$,$d_4$,${s_3})$= $\max(17,22)$ = $22$.
\end{example}
\begin{definition} [Sequence utility]
\label{sequence utility}
The sequence utility of a sequence $t = \langle {t_1}$, ${t_2}$, ..., ${t_m}\rangle $ in $s = \langle {X_1},{X_2},...,{X_n}\rangle $ is denoted and defined as $v\left( {t,s} \right) = \bigcup\limits_{s'\sim t \wedge s' \subseteq s} {u(s')}$. The utility of $t$ in the dataset $SDB$ is denoted and defined as a utility set: $v\left( {t} \right) = \bigcup\limits_{s \in S} {v(t,s)}$.
\end{definition}
\begin{example} \normalfont
The utility of $t$ = $\langle$ $cb$ $\rangle$ in $s_{1}$ is calculated as $v(t$, $s_{1})$ = $\{u(\langle(c,2)$$(b,1)$$\rangle)$, $u(\langle(c,2)$$(b,3)$$\rangle)$, $u(\langle$$(c,3)$$(b,3)$$ \rangle)\}$ = $\{11$, $21$, $24\}$. The utility of $t$ in $SDB$ is $v(t)$= $\{v(t$,$s_{1})$, $v(t,s_{2})$, $v(t,s_{3})$, $v(t,s_{4})\}$ = $\{11$, $21$, $24$, $16$, $16$, $19$, $13$, $13\}$.
\end{example}
\begin{definition} [Sequence maximum utility]
\label{sequence maximum utility}
Given a sequence $t$, the maximum utility of $t$ in $s$ is denoted and defined as ${u_{\max }}$$(t,s)$ = $\max\{u(t,i,s): \forall i \in s' \wedge s'\sim t \wedge s' \subseteq s \}$. The maximum utility of a sequence $t$ in a q-sequence dataset $SDB$ is denoted and defined as ${u_{\max }}(t)$= $\sum {u_{\max }}(t,s): \forall s \in S\} $.
\end{definition}
\begin{example} \normalfont 
The maximum utility of the sequence $t = \langle cb\rangle $ in the sequence dataset $SDB$ shown in Table \ref{sequencedataset} is ${u_{\max}}(t)$ = $u_{\max}(\langle cb \rangle$, $s_{1})$ $+$ $ u_{\max}(\langle cb \rangle$, $s_{2})$ $+$ $u_{\max}(\langle cb \rangle$, $s_{3})$ $+$ $u_{\max}(\langle cb \rangle$, $s_{4})$ = $24 + 16+ 19+ 13 = 72$.
\end{example}

\begin{definition} [high utility sequential pattern]
\label{husp}
A sequence $t$ is said to be a high utility sequential pattern if ${u_{\max }}(t) \ge \xi $, where $minUtil $ is a given user-specified minimum utility threshold. 
For example, given $minUtil = 154$, the complete set of HUSPs in the sequence dataset $SDB$ (Table \ref{sequencedataset}) is shown in Table \ref{huspset}
\end{definition}
\begin{table}[h]
\renewcommand{\arraystretch}{1.0}
\centering
\caption {The set of HUSPs for $minUtil = 154$}
\begin{tabular}{ |c| c| c| c| }
\hline
HUSP & utility & HUSP & utility \\ 
\hline
\hline
$\langle (cg)\rangle$ & $154$ &	$\langle (cg)(be) \rangle$ & $186$ \\ \hline
$\langle (cg)(abf)(be) \rangle$ & $159$ &	$\langle (g) \rangle$ & $203$\\
\hline
$\langle (cg)(ab)(be)) \rangle$ & $155$ &	$\langle (g)(be)\rangle$ & $168$\\
\hline
\end{tabular}
\label{huspset} 
\end{table}
\begin{definition} [Support of a pattern]
\label{support}
Given a sequence $t$ and the dataset $SDB$ = $\{{s_1},{s_2},...,{s_n}\}$, the support (or absolute support or support.count) of the sequence $t$ in $SDB$ is defined as the number of q-sequences that contain $t$ and is denoted by $supp(t)$. Mathematically, the support of $t$ is defined as $supp(t)$ = $|\{s|s \sim t \wedge s \in SDB\}|$. For example, $supp($$\langle$$(cg)$$\rangle)$ = $|\{s_1$,$s_2$,$s_3$,$s_4\}|$ = 4, $supp($$\langle$ $(cg)$$(be)$ $\rangle)$ = $|\{s_1,s_3,s_4\}|$ = 3.
\end{definition}
\begin{definition} [Frequent high utility sequential patterns]
\label{fhusp}
Given a sequence $t$ and the dataset $SDB$ = $\{{s_1},{s_2},...,{s_n}\}$, $t$ is said to be a frequent high utility sequential pattern (FHUSP) if and only if $t$ is a HUSP and $sup(t) \geq minSup$, for a threshold $minSup$ set by the user.
\end{definition}
\begin{definition} [Closed frequent high utility sequential patterns]
\label{def:chusp}
Given a sequence $t$ and the dataset $SDB$ = $\{{s_1},{s_2},...,{s_n}\}$, $t$ is said to be a closed frequent high utility sequential pattern (CHUSP) if and only if $t$ is a FHUSP and there exists no FHUSP that is a proper super-sequence of $t$ and has the same support. Mathematically, the set of all CHUSPs is defined as:
\begin{equation*}
CHUSP = \{s \in FHUSP|s'\notin FHUSP: s \subseteq s' \wedge supp(s)=supp(s') \}
\end{equation*}
\end{definition}
The goal of CHUSPM is to discover the set of CHUSPs that satisfies the definition \ref{def:chusp}.
For example, given $minUtil=130$, $minSup$=50\%, the set of CHUSPs is shown in Table \ref{setchusp}.
\begin{table}[h]
\centering
\caption{The set of CHUSPs for $minUtil$=130, $minSup$=50\%}
\begin{tabular}{|c|c|c|c|c|c|}
\hline
$sequence$ & $u(t)$ & $supp(t)$ \\ \hline
$\langle (abf)(be) \rangle$ & $133$ & $3$  \\ \hline 
$\langle (ab)(be) \rangle$ & $147$ & $4$  \\ \hline 
$\langle (bceg) \rangle$ & $134$ &$2$  \\ \hline 
$\langle (cg)\rangle$ & $154$ & $4$  \\ \hline 
$\langle (cg)(abf)(be)\rangle$ & $159$ & $2$  \\ \hline 
$\langle (cg)(be)\rangle$ & $186$ & $3$  \\ \hline 
$\langle (c)(abf)(be)\rangle$ & $148$ & $3$  \\ \hline 
$\langle (c)(be)\rangle$ & $138$ & $4$  \\ \hline 
\end{tabular}
\label{setchusp} 
\end{table}
\begin{definition} [ULS: utility list structure]
\label{ULS}
Assume that a sequence $t$ has $k$ ($k>0$) ending q-items $i$ in a q-sequence $s$ where $i_1 < i_2 <{\dots}<i_k$. The ULS of $t$ in $s$ is a list of $k$ elements, where the $\alpha^{th} (1 \le \alpha \le k)$ element in the ULS contains\\
$\begin{cases}
tid:~\textnormal{is the itemset ID of}~ i_{\alpha}~\textnormal{of}~t~\textnormal{in}~s \\ 
acu:~\textnormal{is the maximum utility of}~{i_\alpha}~ \textnormal{in}~t \\ 
link:~\textnormal{is a pointer pointing to either the}~(\alpha+1)^{th}~\textnormal{element or}~{null}
\end{cases}$
\end{definition}

\begin{definition} [UCS: utility chain structure]
\label{ucs}
Given a sequence $t$ and $s$. The $UCS$ of $t$ in $s$ is denoted and defined as\\
$UCS(t,s)$=$\begin{cases}
peuts:~\textnormal{is the prefix extension utility of}~t~\textnormal{in}~s\\ 
ULS:~\textnormal{is the ULS of sequence}~t~\textnormal{in}~s
\end{cases}$
\end{definition}
\begin{definition} [CHUS: node structure]
\label{pusp-structure}
Given a sequence $t$, the $CHUS$ of $t$ in $SDB$ is denoted and defined as\\
$CHUS(t)$= $\begin{cases}
sidSet:~\textnormal{the set of sequence IDs containing}~t~\textnormal{in}~SDB\\ 
ucpSet = \bigcup\limits_{s\in S}(UCS(t,s))
\end{cases}$
\end{definition}
\begin{definition}[Concatenation] \label{concatenation}
Given a sequence $t$, there are two types of concatenation of $t$:\\
$\begin{cases}
I-Extension:~\textnormal{to insert an item into the last itemset of}~t\\
S-Extension:~\textnormal{to add a new 1-itemset at the end of}~t
\end{cases}$\\
\end{definition}
\begin{example} \normalfont
For example, $\langle(acg)\rangle$ and $\langle(ac)(a)\rangle$ is generated by performing an I-Extension and an S-Extension of the sequence $\langle(ac)\rangle$, respectively. 
\end{example}

\begin{definition} [SWU: weighted sequence utilization]
\label{swu}
$SWU$ of a sequence $t$ in $SDB$ is defined as
\begin{center}
$SWU(t)$ = $\sum\limits_{s'\sim t \wedge s' \subseteq s \wedge s \subseteq SDB} {u(s)}$.
\end{center}
\end{definition}
For example, $SWU(\langle a(be) \rangle)$ = $u(s_1)$ + $u(s_2)$ + $u(s_3)$ + $u(s_4)$= $91$ + $96$ + $82$ + $114$ = $383$.
\begin{theorem} [Sequence weighted downward closure property]
\textnormal{Given ${t_1}$ and ${t_2}$, if ${t_2}$ contains ${t_1}$, then $SWU({t_2}) \leq SWU({t_1})$.}
\label{swutheorem}
\end{theorem}
Theorem \ref{swutheorem} can be used to evaluate whether an item is promising~\cite{dinh2018efficient,dinh2017mining,dinh2015novel}. The CHUSP algorithm also uses this theorem to prune all items with an SWU $< minUtil$.
\begin{definition} [PEU: prefix extension utility]
\label{peu}
Given a sequence $t$ and $s$. The $PEU$ of $t$ in $s$ is denoted and defined as
\begin{center}
 $PEU(t,s) = \max\{PEU(t,{i_k},s)$ $:\forall {i_k}$ that is an ending q-item of $t$ in $s\}$
\end{center}
\begin{center}
$PEU(t,{i_k},s)$ = $\begin{cases}
u(t,{i_k},s)+ ru({i_k},s),~\textnormal{if}~ru({i_k},s) > 0, \\ 
0,~\textnormal{otherwise}.
\end{cases}$
\end{center}
The $PEU$ of $t$ in $SDB$ is denoted and defined as
\begin{center}
$PEU(t) = \sum\limits_{s'\sim t \wedge s' \subseteq s \wedge s \subseteq S} {PEU(t,s)}$. 
\end{center}
Given ${t_1}$ and ${t_2}$, if ${t_2}$ contains ${t_1}$ then $u({t_2}) \le PEU({t_1})$.
\end{definition}
\begin{definition} [RSU: reduced sequence utility]
\label{rsu}
Given a sequence $t$ and $s$. The RSU of $t$ in $s$ is denoted and defined as \\
\begin{center}
$RSU(t,s)$ = $\begin{cases}
PEU(t')| {t'} \subseteq {t} \wedge {s_1}\sim t \wedge {s_1} \subseteq s \wedge {s_2}\sim t \wedge {s_2} \subseteq s, \\ 
0,~\textnormal{otherwise}.
\end{cases}$
\end{center}
The RSU of the sequence $t$ in SDB is denoted and defined as:
\begin{center}
$RSU(t) = \sum\limits_{s'\sim t \wedge s' \subseteq s \wedge s \subseteq SDB} {RSU(t,s)}$. 
\end{center}
Given ${t_1}$ and ${t_2}$, if ${t_2}$ contains ${t_1}$ then $u({t_2}) \le RSU({t_1})$.
\end{definition}
\begin{theorem} [Pruning strategy by PEU and RSU \cite{wang2016efficiently}]
\textnormal{
Given a pattern $t$, $PEU(t)$ and $RSU({t})$ are considered as upper bounds on the utility of $t$ and its descendants. If $PEU({t}) < minUtil$ or $RSU({t}) < minUtil$, then $t$ and its descendants can be pruned from the search space without affecting the result of the mining process.}
\end{theorem}
\begin{theorem} [MSP: minimum support-based pruning]
\textnormal{
Given a sequence $t$, if $supp(t) < minSup$, then the sequence $t$ and its descendants are not CHUSP.}
\label{msptheorem}
\end{theorem}
\section{The proposed CHUSP algorithm}
\begin{algorithm}[!htb]
\LinesNumbered
\SetKwInOut{Input}{input}\SetKwInOut{Output}{output}
 \Input{$SDB$: a q-sequence dataset, $t$: a sequence with its CHUS, $minUtil$, $minSup$}
 \Output{$CHUSP\_Set$: The set of CHUSPs}
 \BlankLine
 $CHUSP\_Set \leftarrow \emptyset $ \\
 $\neg CHUSP\_Set \leftarrow \emptyset $ \\
 Scan $SDB$ to calculate $SWU$ for all items\\
 Remove all items that have $SWU<minUtil$\\
 \If {$(PEU(t) < minUtil)$}{return}
 Scan the projected dataset to:
 {
 \begin{itemize}
		\item[a.] put I-Extension items into iExts, 
 		\item[b.] put S-Extension items into sExts
 \end{itemize}
 }
 Remove low $RSU$ items from iExts and sExts\\
 \ForEach{ item $i \in$ iExts}
 {
 	$(t',v(t')) \leftarrow$ I-Extension$\left( {t,i} \right)$\\
 	Construct the $CHUS$ structure of $t'$\\
 \If{$ (supp(t') \geq minSup)$}
 {
  		\If{$({u_{\max }}(t') \ge minUtil)$}
   {
  	 		\textnormal{checkClosedPatterns}$(t, t')$\\
   }
  		\textnormal{CHUSP}$(t'$, $minUtil$, $minSup$)
 }
 }
 \ForEach{ item $i \in$ sExts}
 {
 	$(t',v(t')) \leftarrow$ S-Extension$\left( {t,i} \right)$\\
 	Construct the $CHUS$ structure of $t'$\\
 \If{$ (supp(t') \geq minSup)$}
 {
 		  \If{$({u_{\max }}(t') \ge minUtil)$}
    {
 	 		  \textnormal{checkClosedPatterns}$(t, t')$\\
     }
 		  \textnormal{CHUSP}$(t'$, $minUtil$, $minSup$)
 }
 }
 Remove non-CHUSPs from $CHUSP\_Set$ \\
\Return $CHUSP\_Set$;
 \caption{The CHUSP algorithm }
 \label{CHUSP}
\end{algorithm}
\begin{algorithm}[!htb]
\LinesNumbered
\SetKwInOut{Input}{input}\SetKwInOut{Output}{output}
\Input{previous pattern $t$, current pattern $t'$ ($t \subseteq t'$), $CHUSP\_Set$, $\neg CHUSP\_Set$}
\BlankLine
\tcc{In this case: $t$ is not a CHUSP, $t'$ is a candidate}
\If {$supp(t)$==$supp(t')$}
{
 Remove $t$ from $CHUSP\_Set$\\
 Add $t$ into $\neg CHUSP\_Set$\\
 Add $t'$ into $CHUSP\_Set$\\
}
\tcc{In this case: both $t$ and $t'$ are candidates}
\Else
{
 Add $t'$ into $CHUSP\_Set$\\
 \If {$t \notin \neg CHUSP\_Set$}
 {
  Add $t$ into $CHUSP\_Set$\\
 }
}
\caption{\textnormal{checkClosedPatterns} procedure}
\label{check Closed Procedure}
\end{algorithm}
The pseudo-code of the CHUSP algorithm is shown in Algorithm \ref{CHUSP}. The input is a q-sequence dataset $SDB$, a sequence $t$ with its $CHUS$ structure, and three predefined parameters: $minUtil$, $minSup$. First, a set called $CHUSP\_Set$ is initialized to keep all CHUSPs. We also use the $\neg CHUSP\_Set$ to track all but not closed high utility sequential patterns. The algorithm scans $SDB$ to calculate the SWU of all items in $SDB$ (line 3). It then selects all items with an SWU of greater than $minUtil$ and builds the initial CHUS structure and the lexicographic tree required by the mining process. It also removes all items with an SWU value less than $minUtil$ (line 4). 
The topmost node in that tree is the root node, where its children are q-sequences that contain a single item. 
Each node other than the root stores a sequence $t$, the $CHUS$ structure of $t$, utility matrices, remaining utility matrices, and the list that contains sequence IDs called $seqIdList$ of 1 q-items in q-sequences of $SDB$.
If $PEU(t)$ is less than $minUtil$, then
the algorithm will consider $t$ as a leaf and will not expand the lexicographic tree using node $t$, i.e., all its descendants will be pruned (lines 5-6).

In the next step, the algorithm scans the projected dataset that includes the $CHUS$ of $t$ in $SDB$ to collect all items that can be combined with $t$ to form a new sequence by I-Extension or S-Extension (line 7). 
Each item with an RSU value lower than $minUtil$ is discarded from the mining process (line 8).
Then, the algorithm performs a loop over all items in the iExts (lines 9-16) and sExts (lines 17-24).
For each item $i$ in the iExts, the algorithm performs an I-Extension with this item to form a new sequence $t'$ by inserting $i$ in the last itemset of $t$.
In addition, the $CHUS$ structure, $seqIdList$, and the maximum utility of $t'$ are constructed and calculated by extending the $CHUS$ of $t$ (lines 10-11). 
To reduce the search space and enhance the mining process, CHUSP applies the MSP strategy (Theorem \ref{msptheorem}) to discard non-frequent patterns (line 12).
If the condition returns true, CHUSP stops considering these patterns and backtracks to the previous step. Otherwise, the algorithm checks if the pattern's utility value is greater than $minUtil$. If yes, the pattern is a high utility sequential pattern (line 13). CHUSP calls the $checkClosedPatterns$ procedure to check if that HUSP is closed (line 14).

The inputs of $checkClosedPatterns$ procedure are two patterns $t$, $t'$, $CHUSP\_Set$ and $\neg CHUSP\_Set$. Note that the sequence $t'$ is a super-sequence of $t$ by performing the $I-Extension$ or $S-Extension$ concatenation. We consider $t$ and $t'$ as the previous and current sequences since $t'$ is generated from $t$.
The procedure checks if the previous sequence $t$ is a CHUSP by comparing its support count with the support count of the current sequence $t'$. If the two support values equal, it means that $t$ is not a CHUSP because it has a super-sequence with the same support (break the Def.~\ref{def:chusp}), then the procedure checks if $t$ is in $CHUSP\_Set$, if yes then removes $t$ from this set (line 3). The procedure also checks if $t$ is in $\neg CHUSP\_Set$; if No, add $t$ into this set (line 4). The purpose of this action is to track all non-candidate sequences. During mining, $t$ may be extended to other $t'$ by doing other concatenations. In this case, $t$ involves in other checking procedures. The procedure then inserts the current sequence $t'$ into the $CHUSP\_Set$ (line 5). It is worth noting that CHUSP is a recursive algorithm. Thus the current sequence $t'$ will be later called in other rounds of the algorithm to extend itself. In other words, the sequence $t'$ is the super-sequence of a sequence $t$ at this stage, but it will be the sub-sequence of another sequence in another stage. Thus, any sequences in the $CHUSP\_Set$ are candidates and may be removed from the set when the algorithm detects super-sequences having the same support. If the supports of $t$ and $t'$ are different, the two patterns become candidates. The procedure adds the current pattern $t'$ to the $CHUSP\_Set$ as a candidate (line 8). Next, the procedure checks if the previous sequence $t$ is in the $\neg CHUSP\_Set$. If yes, then it will not be a CHUSP candidate. Otherwise, $t$ is inserted into $CHUSP\_Set$. 

The CHUSP recursively calls itself to expand $t'$ (line 15). A similar process is performed for all items in sExts. It passes a sequence and its projected dataset to each recursive call as input parameters. The sequence dataset $SDB$ and lines 1-4 are used only for initializing the algorithm and are not performed during recursive calls. For each item in sExts, a new pattern is generated by performing an {S-Extension} (lines 16 to 22). When the algorithm completes recursive calls, the algorithm traverses all patterns in the $CHUSP\_Set$ to remove non-CHUSPs from this list (line 23). Finally, it returns all CHUSPs as the output.
\section{Comparative experiment}
Experiments were performed to evaluate the performance of CHUSP on a computer with a 64-bit Intel(R) Xeon(R) Gold 6330 CPU @ 2.00GHz, 12 GB of RAM, running Windows 10 Enterprise LTSC. The source code is publicly available on \href{https://github.com/ClarkDinh/CHUSP}{Github}. All the algorithms were implemented in C\#. 
The proposed algorithm was compared with two algorithms. The first algorithm is the HUS-Span algorithm \cite{wang2016efficiently} for mining HUSPs. The second algorithm is FHUSP, an extension of HUS-Span for mining FHUSPs.
The performance of the three algorithms has been compared on real datasets previously used in \cite{dinh2018efficient,le2018pure}.
The characteristics of these datasets are shown in Table \ref{dataset}. They are eight real-life datasets. They have varied characteristics, such as sparse and dense datasets; short and long sequences.
For each dataset, the $minUtil$ was decreased until a clear winner was observed or algorithms became too long to execute. 
In some cases, a constraint on the maximum length of CHUSP ($maxLength$) was used to speed up the experiments.
For $minSup$, a suitable empirical value was chosen for each dataset to ensure that the algorithms discovered a certain number of CHUSPs. The $minSup$ values for Sign, Kosarak10k, BMSWebView1, BMSWebView2, Fifa and Bible were set to $50\%$, $5\%$, $20\%$, $20\%$, $0.5\%$, and $0.5\%$, respectively.

\begin{table} [!htb]
\renewcommand{\arraystretch}{1.0}
\centering
\caption {Characteristics of the datasets} 
\begin{tabular}{|l|c|c|c|c|c|} \hline
Dataset & $\#$Sequence & $\#$Item &Avg. seq length\\ \hline \hline 
Sign &$800$ &$310$ &$51.99$ \\ \hline 
Kosarak10k &$10,000$ &$10,094$ &$8.14$ \\ \hline
BMSWebView1 &$59,601$ &$497$ &$2.51$ \\ \hline
BMSwebview2 &$77,512$ &$3,340$ &$4.62$ \\ \hline
Fifa &$20,450$ &$2,990$ &$34.74$ \\ \hline 
Bible &$36,369$ &$13,905$ &$21.64$ \\ \hline
\end{tabular}
\label{dataset} 
\end{table}


\begin{figure*}[!htb]
\centering 
\includegraphics[width=0.65\linewidth]{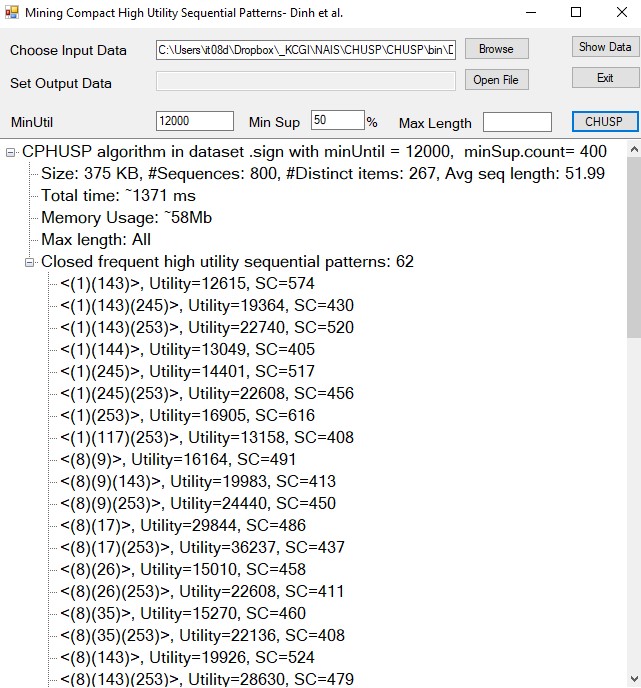} 
\caption{The user interface of the CHUSP application}
\label{Fig:UI}
\end{figure*}

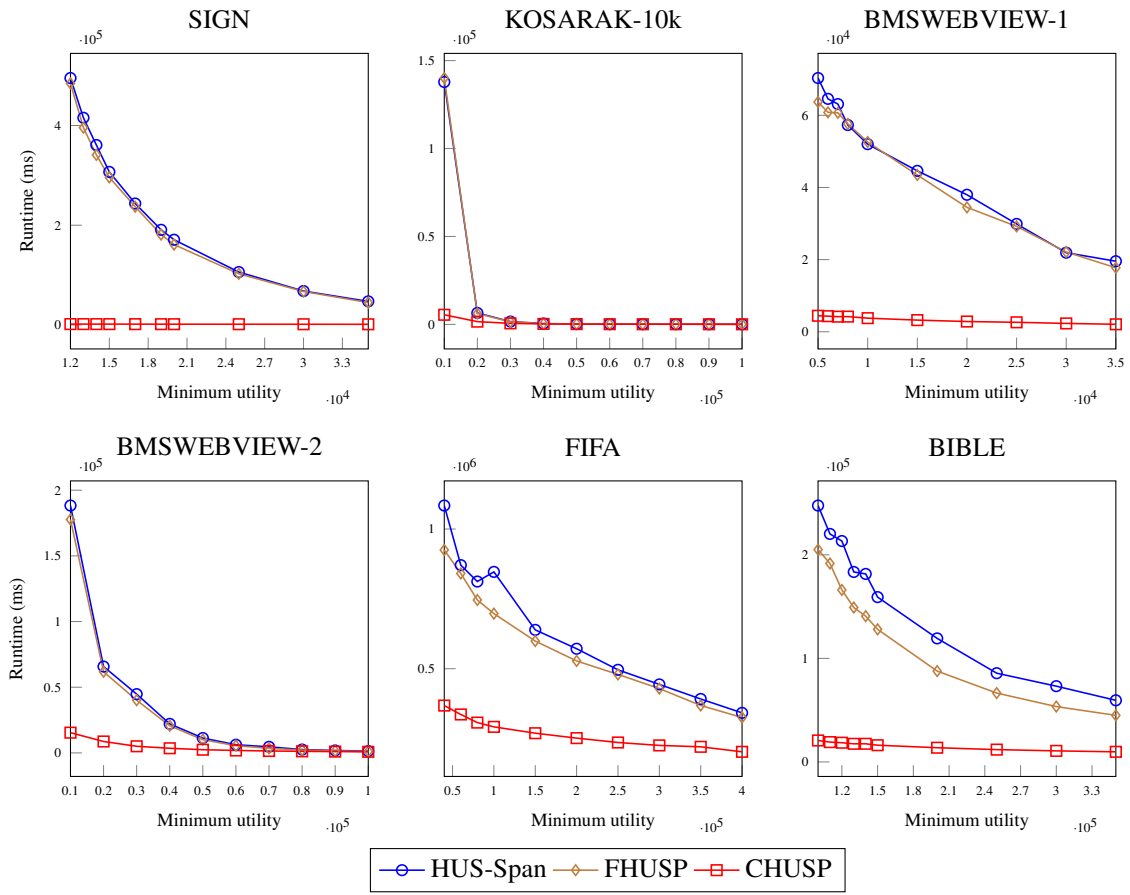
\begin{figure*}[!htb]
\centering
\begin{tikzpicture}
 \begin{groupplot}[group style={group size= 3 by 6},height=5.5cm,width=5.5cm]
 \nextgroupplot[title=SIGN,xlabel={Minimum utility},ylabel={Runtime (ms)}, xmin=12000, xmax=35000, xtick distance= 3000, tick pos=left,ylabel near ticks,xlabel near ticks,label style={font=\tiny\scriptsize},tick label style={font=\tiny},legend style={font=\tiny}, title style={font=\normalsize}, every axis plot/.append style={semithick}]
 \addplot[mark=o,color=blue] 
 plot coordinates {(35000,47236)	(30000,67965)	(25000,106051)	(20000,170870)	(19000,190787)	(17000,243495)	(15000,306935)	(14000,360823)	(13000,415523)	(12000,495239)};
 \addplot[mark=diamond,color=brown] 
 plot coordinates {(35000,44920)	(30000,67075)	(25000,102534)	(20000,161142)	(19000,181166)	(17000,237160)	(15000,295960)	(14000,341095)	(13000,395695)	(12000,484455)};
 \addplot[mark=square,color=red] 
 plot coordinates {(35000,1260)	(30000,1335)	(25000,1400)	(20000,1470)	(19000,1515)	(17000,1555)	(15000,1625)	(14000,1620)	(13000,1660)	(12000,1675)
};

 \coordinate (c1) at (rel axis cs:0,1);
 \nextgroupplot[title=KOSARAK-10k,xlabel={Minimum utility},xmin=10000, xmax=100000,xtick distance=10000,tick pos=left,ylabel near ticks,xlabel near ticks,label style={font=\tiny\scriptsize},tick label style={font=\tiny},legend style={font=\tiny}, title style={font=\normalsize}, every axis plot/.append style={semithick}]
\addplot[mark=o,color=blue] 
 plot coordinates {(100000,75)	(90000,90)	(80000,110)	(70000,120)	(60000,155)	(50000,210)	(40000,475)	(30000,1610)	(20000,6500)	(10000,137834)};
 \addplot[mark=diamond,color=brown]
 plot coordinates {(100000,55)	(90000,85)	(80000,85)	(70000,95)	(60000,130)	(50000,195)	(40000,395)	(30000,1445)	(20000,6200)	(10000,140103)};
 \addplot[mark=square,color=red] 
 plot coordinates {(100000,55)	(90000,80)	(80000,85)	(70000,95)	(60000,120)	(50000,160)	(40000,270)	(30000,580)	(20000,1570)	(10000,5465)};
 \coordinate (c2) at (rel axis cs:1,1);


 \nextgroupplot[title=BMSWEBVIEW-1, xlabel={Minimum utility}, xmin=5000, xmax=35000,xtick distance= 5000,tick pos=left, ylabel near ticks,xlabel near ticks,label style={font=\tiny\scriptsize},tick label style={font=\tiny},legend style={font=\tiny}, title style={font=\normalsize}, every axis plot/.append style={semithick}]
 \addplot[mark=o,color=blue] 
 plot coordinates {(35000,19555)	(30000,21921)	(25000,29895)	(20000,37980)	(15000,44600)	(10000,52007)	(8000,57310)	(7000,63115)	(6000,64570)	(5000,70341)};
 \addplot[mark=diamond,color=brown]
 plot coordinates {(35000,17745)	(30000,22122)	(25000,29215)	(20000,34480)	(15000,43460)	(10000,52597)	(8000,57630)	(7000,60695)	(6000,60820)	(5000,63701)};
\addplot[mark=square,color=red] 
 plot coordinates {(35000,2070)	(30000,2315)	(25000,2625)	(20000,2840)	(15000,3235)	(10000,3775)	(8000,4205)	(7000,4185)	(6000,4320)	(5000,4440)};

 
 \nextgroupplot[title=BMSWEBVIEW-2, xlabel={Minimum utility}, ylabel={Runtime (ms)}, xlabel={Minimum utility}, xmin=10000, xmax=100000,xtick distance= 10000,tick pos=left,ylabel near ticks, xlabel near ticks,label style={font=\tiny\scriptsize},tick label style={font=\tiny},legend style={font=\tiny}, title style={font=\normalsize}, legend to name=runtime1, every axis plot/.append style={semithick}, yshift=-0.75cm]
\addplot[mark=o,color=blue] 
 plot coordinates {(100000,1365)	(90000,1855)	(80000,2542)	(70000,4620)	(60000,6225)	(50000,11305)	(40000,22090)	(30000,44736)	(20000,65752)	(10000,188322)};
 \addplot[mark=diamond,color=brown]
 plot coordinates {(100000,1085)	(90000,1450)	(80000,2170)	(70000,3425)	(60000,5485)	(50000,10035)	(40000,20636)	(30000,40115)	(20000,61780)	(10000,177690)};
\addplot[mark=square,color=red] 
 plot coordinates {(100000,825)	(90000,990)	(80000,1185)	(70000,1460)	(60000,1855)	(50000,2430)	(40000,3530)	(30000,5040)	(20000,8685)	(10000,15410)};
 									
 
 \nextgroupplot[title=FIFA, xlabel={Minimum utility}, xmin=40000, xmax=400000,xtick distance= 50000,tick pos=left,ylabel near ticks,xlabel near ticks,label style={font=\tiny\scriptsize},tick label style={font=\tiny},legend style={font=\tiny}, title style={font=\normalsize}, legend to name=runtime2, every axis plot/.append style={semithick}, yshift=-0.75cm]
\addplot[mark=o,color=blue] 
 plot coordinates { (400000,341890)	(350000,391638)	(300000,443743)	(250000,495977)	(200000,571491)	(150000,639001)	(100000,846624)	(80000,812076)	(60000,870815)	(40000,1084148)};
 \addplot[mark=diamond,color=brown]
 plot coordinates {(400000,325320)	(350000,368375)	(300000,428920)	(250000,479570)	(200000,527731)	(150000,598996)	(100000,697215)	(80000,746050)	(60000,840930)	(40000,925066)};
 \addplot[mark=square,color=red] 
 plot coordinates {(400000,202490)	(350000,220435)	(300000,225396)	(250000,235841)	(200000,251490)	(150000,269086)	(100000,291832)	(80000,307089)	(60000,336195)	(40000,368010)};

 
 \nextgroupplot[title=BIBLE, xlabel={Minimum utility}, xmin=100000, xmax=350000, xtick distance= 30000,tick pos=left,ylabel near ticks, xlabel near ticks, label style={font=\tiny\scriptsize},tick label style={font=\tiny},legend style={font=\tiny}, title style={font=\normalsize},legend style={at={($(0,0)+(1cm,1cm)$)},legend columns=5,fill=none,draw=black,anchor=center,align=center},
 legend to name=runtime3, every axis plot/.append style={semithick}, yshift=-0.75cm]
\addplot[mark=o,color=blue] 
plot coordinates {(350000,59520)	(300000,73175)	(250000,85700)	(200000,119316)	(150000,159175)	(140000,181556)	(130000,183460)	(120000,213321)	(110000,220177)	(100000,247587)};
\addplot[mark=diamond,color=brown]
 plot coordinates {(350000,44995)	(300000,53469)	(250000,66520)	(200000,87733)	(150000,128040)	(140000,140761)	(130000,149087)	(120000,166100)	(110000,191626)	(100000,205061)};
\addplot[mark=square,color=red] 
  plot coordinates {(350000,9780)	(300000,10705)	(250000,11915)	(200000,13750)	(150000,16165)	(140000,17540)	(130000,17510)	(120000,18580)	(110000,19165)	(100000,20740)};			
 \addlegendentry{\normalsize HUS-Span};
 \addlegendentry{\normalsize FHUSP}; 
 \addlegendentry{\normalsize CHUSP};
 \end{groupplot}
 \coordinate (c3) at ($(c1)!.8!(c2)$);
 \node[below] at (c3 |- current bounding box.south)
 {\pgfplotslegendfromname{runtime3}};
\end{tikzpicture}
 \caption{Runtimes for various minimum utility threshold values}
 \label{Fig:executiontime}
\end{figure*}

First, the execution time of CHUSP is compared with HUS-Span and FHUSP. Figure \ref{Fig:executiontime} show that CHUSP outperforms the compared algorithms on all datasets. Each subfigure's vertical and horizontal axes represent the execution time (milliseconds) and minimum utility threshold values, respectively. In general, for all datasets, when the minimum utility threshold is decreased or when datasets contain more sequences or longer sequences, the running time of the algorithms increase. In that case, CHUSP can be much more efficient than the two algorithms, especially on Sign, Bible, BMSWebview1, and FIFA datasets. On Sign ($minSup$=$50\%$) CHUSP is respectively up to $295.7$, $250.3$, $222.7$, $188.9$, $156.6$, $125.9$, $116.2$, $75.8$, $50.9$, and $37.5$ times faster than HUS-Span for $minUtil$ from $12,000$ to $35,000$. It is respectively up to $292.72$, $245.01$, $215.88$, $188.51$, $154.50$, $118.80$, $111.47$, $73.24$, $50.24$, and $35.94$ times faster than FHUSP.
On BMSWebView2 ($minSup$=$50\%$) CHUSP is respectively up to $12.2$, $8.9$, $7.6$, $6.3$, $4.7$, $3.4$, $3.2$, $2.1$, $1.9$, and $1.7$ times faster than HUS-Span for $minUtil$ from $10,000$ to $100,000$. It is respectively up to $11.5$, $8$, $7.1$, $5.8$, $4.1$, $3$, $2.3$, $1.8$, $1.5$, and $1.3$ times faster than FHUSP. Similar results can be observed for other datasets.
The results indicate that the MSP pruning strategy of CHUSP is effective and can prune many non-frequent patterns. In addition, the CHUS structure and pruning strategies are suitable for mining CHUSPs. Thus, the algorithm can facilitate the mining process and prune more non-candidates than HUS-Span and FHUSP algorithms.
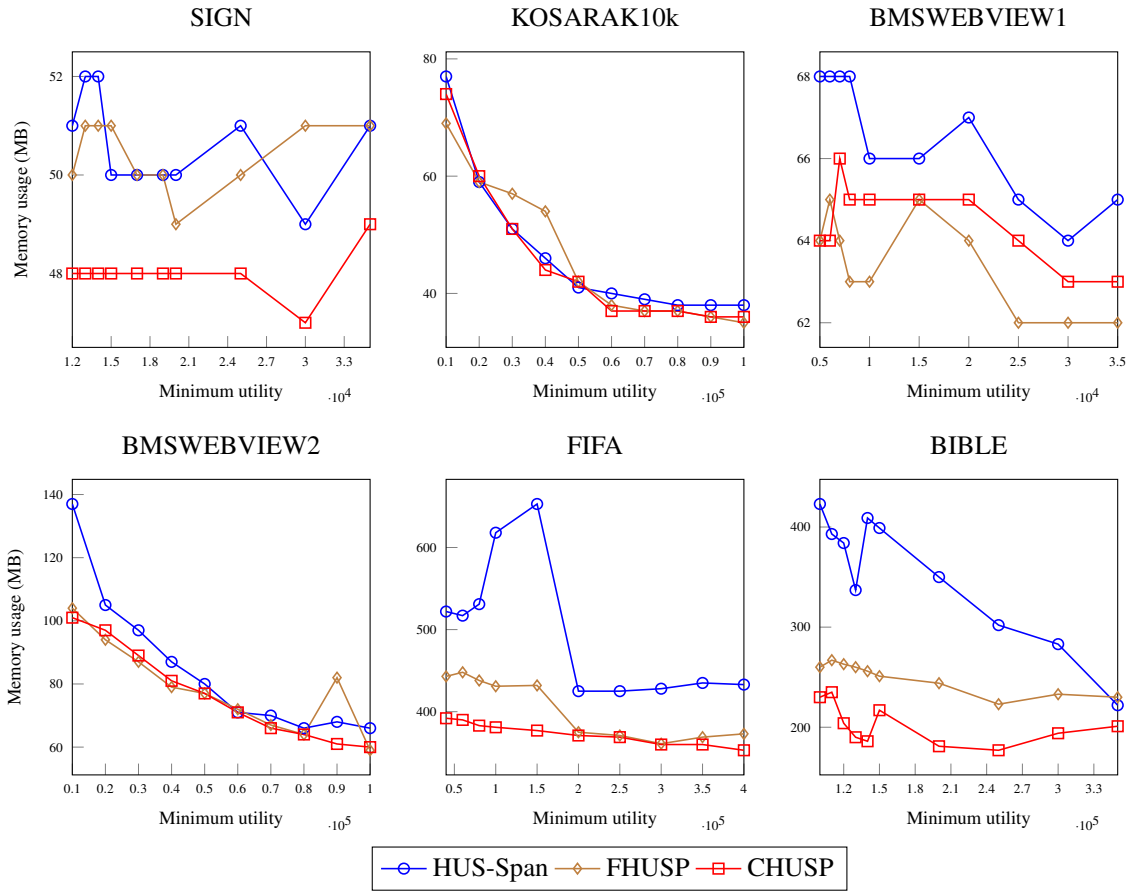
\begin{figure*}[!htb]
\centering
\begin{tikzpicture}
 \begin{groupplot}[group style={group size= 3 by 6},height=5.5cm,width=5.5cm]
 \nextgroupplot[title=SIGN,xlabel={Minimum utility}, ylabel={Memory usage (MB)}, xmin=12000, xmax=35000, xtick distance= 3000, tick pos=left,ylabel near ticks,xlabel near ticks,label style={font=\tiny\scriptsize},tick label style={font=\tiny},legend style={font=\tiny}, title style={font=\normalsize}, every axis plot/.append style={semithick}]
 \addplot[mark=o,color=blue] 
 plot coordinates {(35000,51)	(30000,49)	(25000,51)	(20000,50)	(19000,50)	(17000,50)	(15000,50)	(14000,52)	(13000,52)	(12000,51)};
 \addplot[mark=diamond,color=brown] 
 plot coordinates {(35000,51)	(30000,51)	(25000,50)	(20000,49)	(19000,50)	(17000,50)	(15000,51)	(14000,51)	(13000,51)	(12000,50)};
 \addplot[mark=square,color=red] 
 plot coordinates {(35000,49)	(30000,47)	(25000,48)	(20000,48)	(19000,48)	(17000,48)	(15000,48)	(14000,48)	(13000,48)	(12000,48)
};

 \coordinate (c4) at (rel axis cs:0,1);
 \nextgroupplot[title=KOSARAK10k,xlabel={Minimum utility}, xmin=10000, xmax=100000,xtick distance=10000,tick pos=left,ylabel near ticks,xlabel near ticks,label style={font=\tiny\scriptsize},tick label style={font=\tiny},legend style={font=\tiny}, title style={font=\normalsize}, every axis plot/.append style={semithick}]
\addplot[mark=o,color=blue] 
 plot coordinates {(100000,38)	(90000,38)	(80000,38)	(70000,39)	(60000,40)	(50000,41)	(40000,46)	(30000,51)	(20000,59)	(10000,77)};
 \addplot[mark=diamond,color=brown]
 plot coordinates {(100000,35)	(90000,36)	(80000,37)	(70000,37)	(60000,38)	(50000,42)	(40000,54)	(30000,57)	(20000,59)	(10000,69)};
 \addplot[mark=square,color=red] 
 plot coordinates {(100000,36)	(90000,36)	(80000,37)	(70000,37)	(60000,37)	(50000,42)	(40000,44)	(30000,51)	(20000,60)	(10000,74)};
 \coordinate (c5) at (rel axis cs:1,1);


 \nextgroupplot[title=BMSWEBVIEW1, xlabel={Minimum utility}, xmin=5000, xmax=35000,xtick distance= 5000,tick pos=left, ylabel near ticks,xlabel near ticks,label style={font=\tiny\scriptsize},tick label style={font=\tiny},legend style={font=\tiny}, title style={font=\normalsize}, every axis plot/.append style={semithick}]
 \addplot[mark=o,color=blue] 
 plot coordinates {(35000,65)	(30000,64)	(25000,65)	(20000,67)	(15000,66)	(10000,66)	(8000,68)	(7000,68)	(6000,68)	(5000,68)};
 \addplot[mark=diamond,color=brown]
 plot coordinates {(35000,62)	(30000,62)	(25000,62)	(20000,64)	(15000,65)	(10000,63)	(8000,63)	(7000,64)	(6000,65)	(5000,64)};
\addplot[mark=square,color=red] 
 plot coordinates {(35000,63)	(30000,63)	(25000,64)	(20000,65)	(15000,65)	(10000,65)	(8000,65)	(7000,66)	(6000,64)	(5000,64)};

 
 \nextgroupplot[title=BMSWEBVIEW2, xlabel={Minimum utility}, ylabel={Memory usage (MB)}, xlabel={Minimum utility}, xmin=10000, xmax=100000,xtick distance= 10000,tick pos=left,ylabel near ticks, xlabel near ticks,label style={font=\tiny\scriptsize},tick label style={font=\tiny},legend style={font=\tiny}, title style={font=\normalsize}, legend to name=memory1, every axis plot/.append style={semithick}, yshift=-0.75cm]
\addplot[mark=o,color=blue] 
 plot coordinates {(100000,66)	(90000,68)	(80000,66)	(70000,70)	(60000,71)	(50000,80)	(40000,87)	(30000,97)	(20000,105)	(10000,137)};
 \addplot[mark=diamond,color=brown]
 plot coordinates {(100000,59)	(90000,82)	(80000,64)	(70000,67)	(60000,72)	(50000,77)	(40000,79)	(30000,87)	(20000,94)	(10000,104)};
\addplot[mark=square,color=red] 
 plot coordinates {(100000,60)	(90000,61)	(80000,64)	(70000,66)	(60000,71)	(50000,77)	(40000,81)	(30000,89)	(20000,97)	(10000,101)};

 
 \nextgroupplot[title=FIFA, xlabel={Minimum utility}, xmin=40000, xmax=400000,xtick distance= 50000,tick pos=left,ylabel near ticks,xlabel near ticks,label style={font=\tiny\scriptsize},tick label style={font=\tiny},legend style={font=\tiny}, title style={font=\normalsize}, legend to name=memory2, every axis plot/.append style={semithick}, yshift=-0.75cm]
\addplot[mark=o,color=blue] 
 plot coordinates {(400000,433)	(350000,435)	(300000,428)	(250000,425)	(200000,425)	(150000,653)	(100000,618)	(80000,531)	(60000,517)	(40000,522)};
 \addplot[mark=diamond,color=brown]
 plot coordinates {(400000,373)	(350000,369)	(300000,361)	(250000,371)	(200000,375)	(150000,432)	(100000,431)	(80000,438)	(60000,448)	(40000,443)};
 \addplot[mark=square,color=red] 
 plot coordinates {(400000,353)	(350000,360)	(300000,360)	(250000,369)	(200000,371)	(150000,377)	(100000,381)	(80000,383)	(60000,390)	(40000,392)};

 
 \nextgroupplot[title=BIBLE, xlabel={Minimum utility}, xmin=100000, xmax=350000, xtick distance= 30000,tick pos=left,ylabel near ticks, xlabel near ticks, label style={font=\tiny\scriptsize},tick label style={font=\tiny},legend style={font=\tiny}, title style={font=\normalsize},legend style={at={($(0,0)+(1cm,1cm)$)},legend columns=5,fill=none,draw=black,anchor=center,align=center},
 legend to name=memory3, every axis plot/.append style={semithick}, yshift=-0.75cm]
\addplot[mark=o,color=blue] 
plot coordinates {(350000,222)	(300000,283)	(250000,302)	(200000,350)	(150000,399)	(140000,409)	(130000,337)	(120000,384)	(110000,393)	(100000,423)};
\addplot[mark=diamond,color=brown]
 plot coordinates {(350000,230)	(300000,233)	(250000,223)	(200000,244)	(150000,251)	(140000,256)	(130000,260)	(120000,263)	(110000,267)	(100000,260)};
\addplot[mark=square,color=red] 
  plot coordinates {(350000,201)	(300000,194)	(250000,177)	(200000,181)	(150000,217)	(140000,186)	(130000,190)	(120000,204)	(110000,235)	(100000,230)};

 \addlegendentry{\normalsize HUS-Span};
 \addlegendentry{\normalsize FHUSP}; 
 \addlegendentry{\normalsize CHUSP};
 \end{groupplot}
 \coordinate (c6) at ($(c4)!.8!(c5)$);
 \node[below] at (c6 |- current bounding box.south)
 {\pgfplotslegendfromname{memory3}};
\end{tikzpicture}
\caption{Memory usage for various minimum utility threshold values}
 \label{Fig:memory_usage}
\end{figure*}

Second, the algorithms have also been compared in terms of memory performance for the six datasets for the same $minUtil$, $minSup$, and $maxLength$ values as in the runtime experiment.
Results are shown in Figure \ref{Fig:memory_usage} in terms of memory usage (vertical axes) for various minimum utility values (horizontal axes).
CHUSP consumes less memory than HUS-Span in all cases. It means that the CHUSP structure is more effective than the structure used by the HUS-Span algorithm. In addition, the MSP strategy can filter many non-frequent candidates. CHUSP is also better than FHUSP in most cases, although they are very close in some cases. On FIFA and Bible, we can observe that CHUSP performs much better than FHUSP. Except for the BMSWebview1 dataset, FHUSP consumes less memory than CHUSP on large $minUtil$ values. However, for low $minUtil$ values, when the algorithms need more time to mine patterns, CHUSP outperforms FHUSP. Generally, for each dataset, the memory usage increases when the minimum utility threshold is decreased, and it is also greater for larger datasets. 
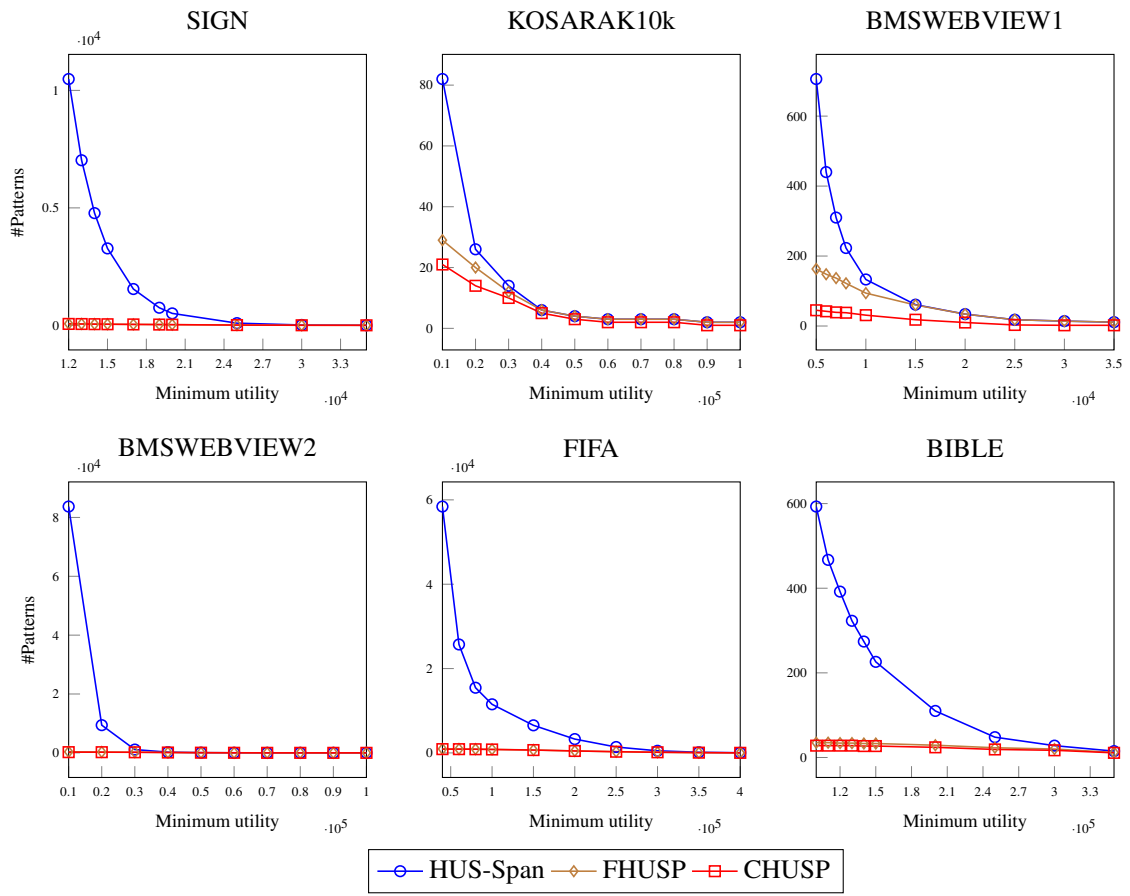
\begin{figure*}[!htb]
\centering
\begin{tikzpicture}
 \begin{groupplot}[group style={group size= 3 by 6},height=5.5cm,width=5.5cm]
 \nextgroupplot[title=SIGN,xlabel={Minimum utility},ylabel={\#Patterns}, xmin=12000, xmax=35000, xtick distance= 3000, tick pos=left,ylabel near ticks,xlabel near ticks,label style={font=\tiny\scriptsize},tick label style={font=\tiny},legend style={font=\tiny}, title style={font=\normalsize}, every axis plot/.append style={semithick}]
 \addplot[mark=o,color=blue] 
 plot coordinates {(35000,4)	(30000,17)	(25000,95)	(20000,510)	(19000,751)	(17000,1547)	(15000,3275)	(14000,4777)	(13000,7027)	(12000,10483)};
 \addplot[mark=diamond,color=brown] 
 plot coordinates {(35000,3)	(30000,5)	(25000,14)	(20000,32)	(19000,36)	(17000,42)	(15000,50)	(14000,56)	(13000,63)	(12000,70)};
 \addplot[mark=square,color=red] 
 plot coordinates {(35000,3)	(30000,5)	(25000,13)	(20000,31)	(19000,35)	(17000,41)	(15000,48)	(14000,51)	(13000,58)	(12000,62)};

 \coordinate (c7) at (rel axis cs:0,1);
 \nextgroupplot[title=KOSARAK10k,xlabel={Minimum utility},xmin=10000, xmax=100000,xtick distance=10000,tick pos=left,ylabel near ticks,xlabel near ticks,label style={font=\tiny\scriptsize},tick label style={font=\tiny},legend style={font=\tiny}, title style={font=\normalsize}, every axis plot/.append style={semithick}]
\addplot[mark=o,color=blue] 
 plot coordinates {(100000,2)	(90000,2)	(80000,3)	(70000,3)	(60000,3)	(50000,4)	(40000,6)	(30000,14)	(20000,26)	(10000,82)};
 \addplot[mark=diamond,color=brown]
 plot coordinates {(100000,2)	(90000,2)	(80000,3)	(70000,3)	(60000,3)	(50000,4)	(40000,6)	(30000,12)	(20000,20)	(10000,29)};
 \addplot[mark=square,color=red] 
 plot coordinates {(100000,1)	(90000,1)	(80000,2)	(70000,2)	(60000,2)	(50000,3)	(40000,5)	(30000,10)	(20000,14)	(10000,21)};
 \coordinate (c8) at (rel axis cs:1,1);


 \nextgroupplot[title=BMSWEBVIEW1, xlabel={Minimum utility}, xmin=5000, xmax=35000,xtick distance= 5000,tick pos=left, ylabel near ticks,xlabel near ticks,label style={font=\tiny\scriptsize},tick label style={font=\tiny},legend style={font=\tiny}, title style={font=\normalsize}, every axis plot/.append style={semithick}]
 \addplot[mark=o,color=blue] 
 plot coordinates { (35000,11)	(30000,14)	(25000,18)	(20000,34)	(15000,61)	(10000,133)	(8000,223)	(7000,310)	(6000,440)	(5000,706)};
 \addplot[mark=diamond,color=brown]
 plot coordinates {(35000,11)	(30000,14)	(25000,18)	(20000,34)	(15000,60)	(10000,94)	(8000,122)	(7000,137)	(6000,148)	(5000,163)};
\addplot[mark=square,color=red] 
 plot coordinates {(35000,2)	(30000,2)	(25000,3)	(20000,10)	(15000,18)	(10000,31)	(8000,38)	(7000,39)	(6000,42)	(5000,45)};


 
 \nextgroupplot[title=BMSWEBVIEW2, xlabel={Minimum utility}, ylabel={\#Patterns}, xlabel={Minimum utility}, xmin=10000, xmax=100000,xtick distance= 10000,tick pos=left,ylabel near ticks, xlabel near ticks,label style={font=\tiny\scriptsize},tick label style={font=\tiny},legend style={font=\tiny}, title style={font=\normalsize}, legend to name=numPattern1, every axis plot/.append style={semithick}, yshift=-0.75cm]
\addplot[mark=o,color=blue] 
 plot coordinates {(100000,1)	(90000,2)	(80000,5)	(70000,8)	(60000,25)	(50000,66)	(40000,239)	(30000,1110)	(20000,9408)	(10000,83699)};
 \addplot[mark=diamond,color=brown]
 plot coordinates {(100000,1)	(90000,2)	(80000,5)	(70000,8)	(60000,25)	(50000,62)	(40000,131)	(30000,217)	(20000,284)	(10000,356)};
\addplot[mark=square,color=red] 
 plot coordinates {(100000,0)	(90000,0)	(80000,1)	(70000,3)	(60000,19)	(50000,50)	(40000,109)	(30000,186)	(20000,220)	(10000,233)};

 
 \nextgroupplot[title=FIFA, xlabel={Minimum utility}, xmin=40000, xmax=400000,xtick distance= 50000,tick pos=left,ylabel near ticks,xlabel near ticks,label style={font=\tiny\scriptsize},tick label style={font=\tiny},legend style={font=\tiny}, title style={font=\normalsize}, legend to name=numPattern2, every axis plot/.append style={semithick}, yshift=-0.75cm]
\addplot[mark=o,color=blue] 
 plot coordinates {(400000,43)	(350000,160)	(300000,504)	(250000,1397)	(200000,3280)	(150000,6543)	(100000,11524)	(80000,15456)	(60000,25715)	(40000,58415)};
 \addplot[mark=diamond,color=brown]
 plot coordinates {(400000,19)	(350000,63)	(300000,155)	(250000,292)	(200000,472)	(150000,694)	(100000,840)	(80000,897)	(60000,925)	(40000,934)};
 \addplot[mark=square,color=red] 
 plot coordinates {(400000,19)	(350000,62)	(300000,151)	(250000,285)	(200000,463)	(150000,676)	(100000,816)	(80000,869)	(60000,890)	(40000,896)};

 
 \nextgroupplot[title=BIBLE, xlabel={Minimum utility}, xmin=100000, xmax=350000, xtick distance= 30000,tick pos=left,ylabel near ticks, xlabel near ticks, label style={font=\tiny\scriptsize},tick label style={font=\tiny},legend style={font=\tiny}, title style={font=\normalsize},legend style={at={($(0,0)+(1cm,1cm)$)},legend columns=5,fill=none,draw=black,anchor=center,align=center},
 legend to name=numPattern3, every axis plot/.append style={semithick}, yshift=-0.75cm]
\addplot[mark=o,color=blue] 
plot coordinates {(350000,15)	(300000,28)	(250000,48)	(200000,110)	(150000,226)	(140000,274)	(130000,323)	(120000,392)	(110000,467)	(100000,593)};
\addplot[mark=diamond,color=brown]
 plot coordinates {(350000,12)	(300000,20)	(250000,23)	(200000,29)	(150000,33)	(140000,33)	(130000,34)	(120000,34)	(110000,35)	(100000,35)};
\addplot[mark=square,color=red] 
  plot coordinates {(350000,11)	(300000,17)	(250000,19)	(200000,24)	(150000,27)	(140000,27)	(130000,28)	(120000,28)	(110000,28)	(100000,28)};

 \addlegendentry{\normalsize HUS-Span};
 \addlegendentry{\normalsize FHUSP}; 
 \addlegendentry{\normalsize CHUSP};
 \end{groupplot}
 \coordinate (c9) at ($(c7)!.8!(c8)$);
 \node[below] at (c3 |- current bounding box.south)
 {\pgfplotslegendfromname{numPattern3}};
\end{tikzpicture}
 \caption{Number of patterns for various minimum utility threshold values}
 \label{Fig:numberpatterns}
\end{figure*}

Finally, the number of patterns was measured for various $minUtil$ threshold values on each dataset. In Figure \ref{Fig:numberpatterns}, vertical axes denote the number of patterns, and horizontal axes indicate the corresponding maximum threshold values. The number of patterns generated by CHUSP is much less than that of HUS-Span and FHUSP for each dataset. On Sign ($minSup$=$50\%$), for $minUtil$ from 12,000 to 35,000, CHUSP found $62$, $58$, $51$, $48$, $41$, $35$, $31$, $13$, $5$, and $3$, respectively. It can be observed that the number of patterns by CHUSP was respectively up to $169.1$, $121.2$, $93.7$, $68.2$, $37.7$, $21.5$, $16.5$, $7.3$, $3.4$, and $1.3$ times less than those found by HUS-Span. In addition, the number of patterns by CHUSP was up to $1.13$, $1.09$, $1.1$, $1.04$, $1.02$, $1.03$, $1.03$, $1.08$, $1.00$, and $1.00$ times less than those found by FHUSP.
On Kosarak10k ($minSup$=$5\%$), the $maxLength$ was set to 3 for the $minUtil$ values of $10,000$ and $20,000$ for HUS-Span and FHUSP; for CHUSP, this parameter was set to $full$. For $minUtil$ from $10,000$ to $100,000$, CHUSP found $21$, $14$, $10$, $5$, $3$, $2$, $2$, $2$, $1$, and $1$ CHUSPs, respectively. It can be observed that the number of patterns by CHUSP was respectively up to $3.9$, $1.9$, $1.4$, $1.2$, $1.3$, $1.5$, $1.5$, $1.5$, $2.0$, and $2.0$ times less than those by HUS-Span. In addition, the number of patterns by CHUSP was up to $1.4$, $1.4$, $1.2$, $1.2$, $1.3$, $1.5$, $1.5$, $1.5$, $2.0$, and $2.0$ times less than those by FHUSP.
On BMSwebview1 ($minSup$=$0.5\%$). The $maxLength$ was set to $3$ for HUS-Span and FHUSP; for CHUSP, this parameter was set to $full$. For $minUtil$ from $5,000$ to $35,000$, CHUSP found $45$, $42$, $39$, $38$, $31$, $18$, $10$, $3$, $2$, and $2$ CHUSPs, respectively. It can be observed that the number of patterns by CHUSP was respectively up to $3.6$, $3.5$, $3.5$, $3.2$, $3.0$, $3.3$, $3.4$, $6$, $7$, and $5.5$ times less than those by HUS-Span. In addition, the number of patterns by CHUSP was up to $3.6$, $3.5$, $3.5$, $3.2$, $3.0$, $3.3$, $3.4$, $6$, $7$, and $5.5$ times less than those by FHUSP.  Similar results can be observed for the BMSwebview1, FIFA, and BIBLE datasets.
These results indicate that the CHUSP algorithm can eliminate many non-candidate patterns from the search space and reduce the number of patterns from the mining process.
\section{Conclusion}
This paper proposed an algorithm named CHUSP for mining closed high utility sequential patterns. The proposed algorithm uses the CHUS structure for efficiently mining CHUSP. Experimental results indicate that CHUSP outperforms HUS-Span and FHUSP algorithms in terms of execution time and memory usage. The number of patterns generated by the three algorithms was also measured for various minimum utility threshold values. The results show that all the pruning strategies used in CHUSP can eliminate many non-CHUSP and thus speed up the mining process. In future work, we will design a parallel framework that can enhance the computational cost of CHUSP and extend the pattern mining framework for other tasks \cite{dinh2018efficient,fournier2021discovering,dinh2019estimating,dinh2020k,dinh2021clustering}.

\bibliography{main.bib}
\end{document}